\documentclass{osa-article}

\journal{osajournal}
\articletype{Research Article}
\usepackage[export]{adjustbox}
\usepackage{amsmath}
\usepackage{amssymb}
\usepackage[section]{placeins}

\begin{document}

\title{Integrated microwave acousto-optic frequency shifter on thin-film lithium niobate}

\author{Linbo Shao\authormark{1,5}, Neil Sinclair\authormark{1,2}, James Leatham\authormark{3}, Yaowen Hu\authormark{1}, Mengjie Yu\authormark{1}, Terry Turpin\authormark{4}, Devon Crowe\authormark{3}, Marko Lon{\v c}ar\authormark{1,6}}

\address{\authormark{1}John A. Paulson School of Engineering and Applied Sciences, Harvard University, 29 Oxford Street, Cambridge, MA 02138, USA\\
\authormark{2}Division of Physics, Mathematics and Astronomy, and Alliance for Quantum Technologies (AQT), California Institute of Technology, 1200 E. California Blvd., Pasadena, California 91125, USA\\
\authormark{3}Raytheon Space \& Airborne Systems, 2000 E. El Segundo Ave., El Segundo, CA 90245, USA\\
\authormark{4}49918 Evergreen Ave., Columbia, MD 21046, USA
}

\email{\authormark{5}shaolb@seas.harvard.edu} %% email address is required
\email{\authormark{6}loncar@seas.harvard.edu} %% email address is required

%%%%%%%%%%%%%%%%%%% abstract %%%%%%%%%%%%%%%%
\begin{abstract*}
Electrically driven acousto-optic devices that provide beam deflection and optical frequency shifting have broad applications from pulse synthesis to heterodyne detection. 
Commercially available acousto-optic modulators are based on bulk materials and consume Watts of radio frequency power. 
Here, we demonstrate an integrated 3-GHz acousto-optic frequency shifter on thin-film lithium niobate, featuring a carrier suppression over 30 dB. 
Further, we demonstrate a gigahertz-spaced optical frequency comb featuring more than 200 lines over a 0.6-THz optical bandwidth by recirculating the light in an active frequency shifting loop. 
Our integrated acousto-optic platform leads to the development of on-chip optical routing, isolation, and microwave signal processing.
\end{abstract*}

%%%%%%%%%%%%%%%%%%%%%%%%%%  body  %%%%%%%%%%%%%%%%%%%%%%%%%%
\section{Introduction}
Integrated acousto-optic or Brillouin scattering devices \cite{Eggleton2019NP} have enabled a wide range of applications including frequency shifting \cite{cheng1992apl, Dong2015adp, li2019aplphoton, Liu2019optica}, microwave-to-optical conversion (modulation) \cite{salram2016np, Cai2019PR, Jiang2019Optica, Jiang2019arxiv, liang2017optica, shao2019optica}, microwave photonic filtering \cite{marpaung2015optica}, frequency comb generation \cite{savchenkoc2011ol}, pulse shaping \cite{Fan2019np}, ultra-narrow-linewidth lasing \cite{Gundacarapu2019NP, otterstrom2018science}, and nonreciprocal transmission \cite{Fang2017NP, Kim2015NP, Kittlaus2018NP, Otterstrom2019Optica, Kittlaus2020Arxiv, Sohn2019arxiv, sohn2018np}. 
Such integrated devices employ photoelasticity and optical confinement of thin-film materials such as silicon \cite{Fang2017NP, Kittlaus2018NP, otterstrom2018science, Otterstrom2019Optica, Kittlaus2020Arxiv}, silicon nitride \cite{Gundacarapu2019NP}, aluminum nitride \cite{Fan2019np, li2019aplphoton, Liu2019optica, Sohn2019arxiv, sohn2018np}, gallium arsenide \cite{salram2016np}, arsenic trisulfide \cite{marpaung2015optica}, lithium tantalate \cite{savchenkoc2011ol}, and lithium niobate (LN) \cite{Cai2019PR, Jiang2019Optica, Jiang2019arxiv, liang2017optica, shao2019optica}. 
Acousto-optic frequency shifters (AOFSs) deflect the light into a different spatial mode and shift its optical frequency by the acoustic frequency.
Commercial AOFSs\cite{savage2010np} employ bulk acoustic waves and provide outstanding extinction ratio, large carrier suppression, and high efficiencies for large optical bandwidth. 
However, they are large discrete components and require a few Watts of radio frequency (RF) power. 
For example, a free-space tellurium dioxide acousto-optic modulator (G\&H AOMO 3110-197) operates at 110 MHz, consumes 2.0 W RF power, and deflects >90\% of the input light around 1060 nm; a fiber-pigtailed AOFS (Brimrose IPF-1500-1550-3FP) operates at 1.5 GHz, deflects 3-10\% of the input light to the first order output fiber. 
To develop integrated AOFS, surface acoustic waves have been employed to deflect light confined by an ion diffused layer \cite{cheng1992apl}, but its relatively large optical mode size (a few microns) limits interactions with sub-micron-wavelength gigahertz acoustic waves.
Recently, electromechanically driven suspended acousto-optic waveguides have been utilized to achieve frequency shifts exceeding 10 GHz\cite{li2019aplphoton, Liu2019optica} but suffer from low efficiencies of $\sim10^{-5}$ and weak carrier suppression. 
In additional to AOFSs, electro-optic devices can achieve optical frequency shifting by destructive interference between Mach-Zehnder modulators \cite{Higuma2001EL, ogiso2010IEEEptl}, by serrodyne frequency shifting \cite{Houtz2009OE, Johnson2010OL, Spuesens2016CLEO}, and by employing electro-optic cavities \cite{Preble2007NPhoton, savchenkov2009ol, Yu2018PR}. 

Here, we demonstrate an integrated gigahertz AOFSs on thin-film LN, which leverage the large piezoelectric and photoelastic coefficients of LN, as well as its low microwave and optical propagation loss. 
Benefiting from the fact that both optical and acoustic indices (phase velocities) of LN are greater (smaller) than those of the underlying silicon dioxide, we construct fully supported AOFSs, thereby providing improved robustness and a more straightforward fabrication process than suspended acousto-optic devices. 
Specifically, at the telecommunication wavelength of 1.5 $\mu$m, we demonstrate an optical frequency shift of 3 GHz with carrier suppression over 30 dB and the opposite sideband suppression >40 dB (below the noise floor). 
Furthermore, as an application of our device, we demonstrate a 3-GHz optical frequency comb with more than 200 lines over a 0.6-THz (5 nm) optical bandwidth by recirculating light in the active frequency shifting loop with our AOFS and an erbium-doped fiber amplifier (EDFA).

\section{Device design and fabrication}
\begin{figure*}[hbt]
\centering
\includegraphics[center]{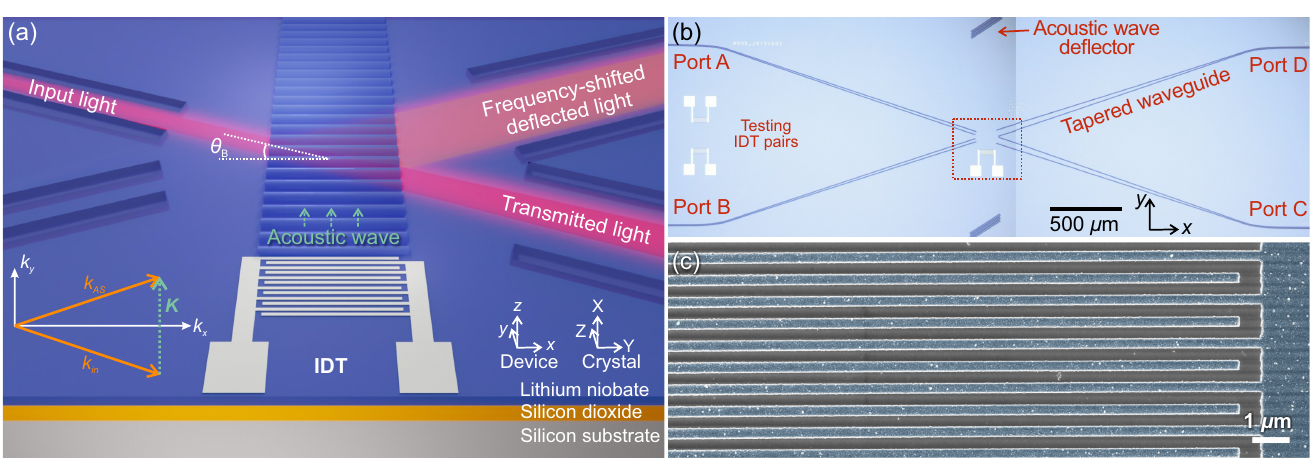}
\caption{ \label{Fig1}
Thin-film lithium niobate (LN) acousto-optic frequency shifter (AOFS).
(a) Schematic of the AOFS. The device is on a LN thin film on a silicon dioxide layer. Input light interacts with the traveling acoustic wave generated by the interdigital transducer (IDT), is partially deflected by the Bragg angle, and is frequency shifted by the acoustic frequency. 
The LN thin film is X cut, the coordinates for the crystal and device are shown. 
The input light, the deflected light, and the acoustic wave satisfy energy and momentum matching conditions (Inset). 
(b) Microscopic image of the fabricated device. Optical waveguides are coupled using lensed fibers and broaden to the acousto-optic region, as indicated by the red dashed lines. A pair of IDTs featuring the same specifications are added for characterizing the microwave-to-acoustic transduction. 
This image is stitched from two microscopic fields of view. 
(c) False-colored scanning electron microscopic image of the IDT. The aluminum region is in blue. }
\end{figure*}

Our integrated AOFSs are fabricated using an 800-nm-thick X-cut LN thin film on a 2-$\mu$m-thick silicon dioxide layer on a silicon substrate (Fig.~\ref{Fig1}(a)). 
The phase matching condition between the light and acoustic wave (Fig.~\ref{Fig1}(a) inset) determines the optimum incident (Bragg) angle $\theta_B$ of the input light beam, given by
\begin{equation}
    \sin\theta_B=\frac{K}{2\ k}=\frac{\lambda}{2\ \Lambda\ n_{\mathrm{eff}}}.    
\end{equation}
The optical wavenumber $k=2\pi n_{\mathrm{eff}}/\lambda$, $n_{\mathrm{eff}}$ is the optical mode index, and $\lambda$ is the optical wavelength in vacuum. 
The acoustic wavenumber $K=2\pi/\Lambda$, where $\Lambda$ is the acoustic wavelength in the LN thin film. 
For the demonstrated 3-GHz AOFS in this work, we use $\Lambda=1.2$ $\mu$m,    $\lambda=1.5$ $\mu$m, and $n_{\mathrm{eff}}=2.0$ yielding $\sin\theta_B=0.3$ ($\theta_B \sim 18$ degrees). 
The maximum shifting frequency of $\sim$9 GHz for $\lambda=1.5$ $\mu$m on the LN thin film occurs when the shifted light is fully reflected by the acoustic grating (in this case, $\theta_B=\pi/2$ or $\sin\theta_B=1$). 

For our device, the light is coupled onto the chip by a lensed fiber and guided by a rib optical waveguide that is defined by two etched grooves on the LN thin film. 
Following the waveguide, the input light bends and adiabatically broadens towards the central acousto-optic region at the Bragg angle (Fig.~\ref{Fig1}(b)).
The rib waveguides terminate near the acousto-optic region, and the input light beam then propagates as a two-dimensional Gaussian beam in the LN thin film. 
The end width of the input waveguides is 18 $\mu$m such that the Rayleigh range of the light beam is much larger than the acoustic wave width of 100 $\mu$m. 
A waveguide collects the expanded light beam deflected by the acoustic waves, then tapers to a on-chip routing waveguide. 
The 40-$\mu$m width of the collecting waveguide is experimentally selected to provide the highest collection efficiency. 
With two optical input waveguides and two output waveguides, our device is configured for either anti-Stokes frequency shifts of the input light at Port A or Stokes frequency shifts of the input light at Port B (Fig.~\ref{Fig1}(b)).

The thin-film acoustic wave propagates along the $y$ direction and is electrically generated by an interdigital transducer (IDT) via piezoelectricity. 
The IDT consists of cross-finger electrodes that are made of a 140-nm-thick aluminum layer deposited by thermal evaporation and patterned by the lift-off process (Fig.~\ref{Fig1}(c)). 
The pitch of the IDT electrodes is 580 nm, which is equal to the half acoustic wavelength (in the IDT region). 
To avoid any acoustic reflection by the chip edge or other devices, etching grooves are fabricated at the far ends to deflect acoustic waves. 
Otherwise, standing acoustic waves formed by any reflection would result in undesired carrier and sideband light at the output.

\section{Acousto-optic interactions}
\begin{figure*}
    \centering
    \includegraphics[center]{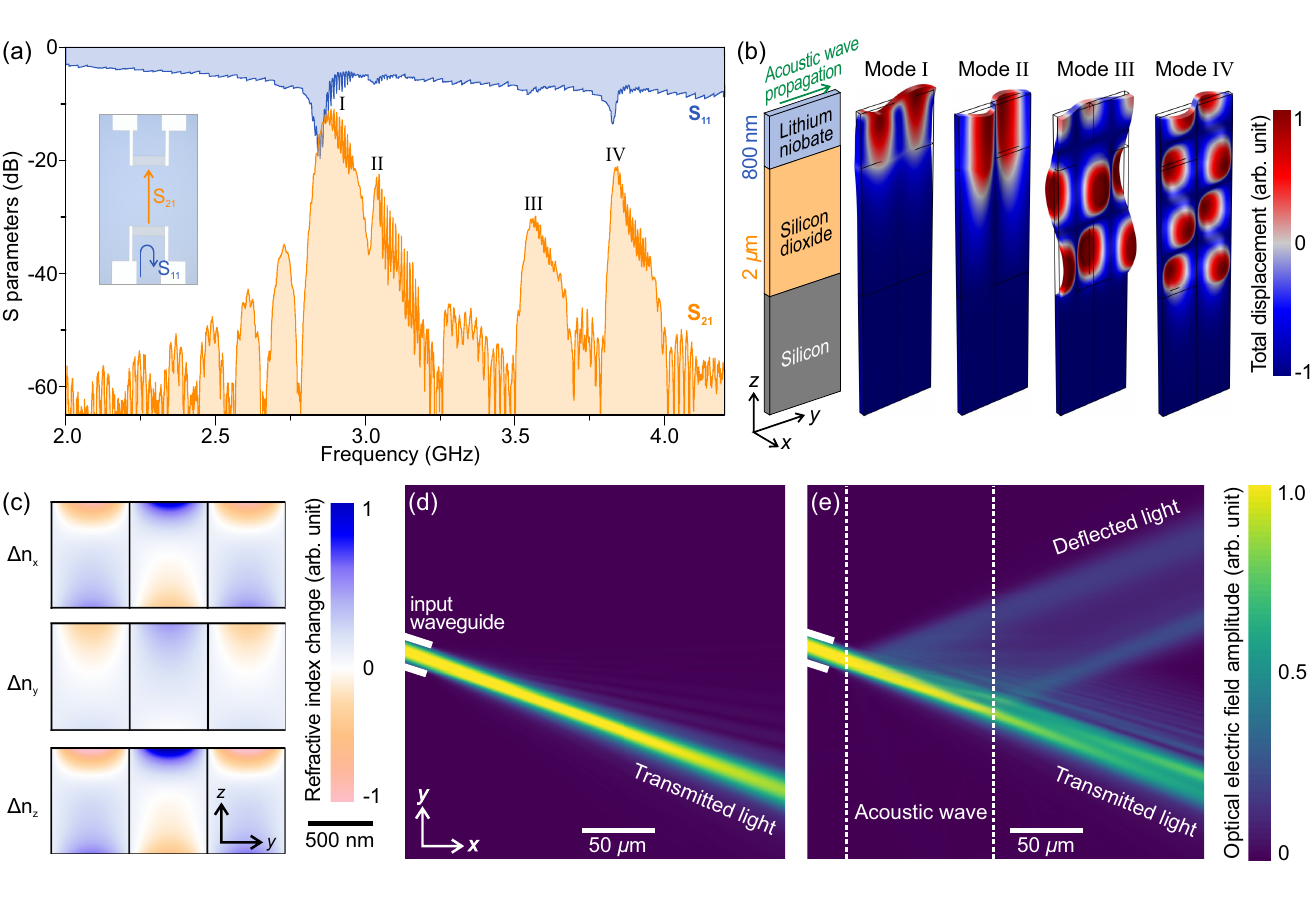}
    \caption{
    Measurements of microwave-to-acoustic transduction and simulations of acousto-optic interactions. 
    (a) Measured reflection $S_{11}$ and transmission $S_{21}$ spectra of an IDT pair. Four acoustic modes (I-IV) are identified in the transmission $S_{21}$ spectrum. 
    (b) Simulated displacement profiles of acoustic modes corresponding to the peaks in the transmission $S_{21}$ spectrum.
    (c) Calculated refractive index variations of the LN thin film induced by the strain fields of acoustic Mode I. The refractive index variations in all directions are normalized using the same scale. 
    (d)(e) Optical electric field profiles from 3D FDTD simulations of the device (d) without and (e) with the acoustic wave. These profiles show the magnitude of optical electric fields at the center of the LN thin film, which is 400 $\mu$m from the top surface.
    All coordinates used here are the device coordinate.}
    \label{Fig2}
\end{figure*}

To characterize the microwave-to-acoustic transduction by the IDT, we measure the microwave reflection and transmission scattering ($S$) parameter spectra of an IDT pair (Fig.~\ref{Fig2}(a)). 
Multiple acoustic modes are observed in the transmission spectra (labeled I-IV in Fig.~\ref{Fig2}(a)), consistent with the acoustic modes found in the numerical simulations (Fig.~\ref{Fig2}(b)). 
They are referred to as the Rayleigh wave (Mode I), Love wave (Mode II) of LN thin film, and higher-order Love waves (Modes III and IV) that partially propagating in the silicon dioxide underlying layer. 
We employ the Rayleigh wave (Mode I) for our AOFS due to its highest transduction efficiency. 
Twenty-five pairs of 100-$\mu$m-width electrodes are used in each IDT to match the external impedance of 50 $\Omega$, as indicated by the over 10 dB dip in the $S_{11}$ spectrum (Fig.~\ref{Fig2}(a)). 
In the case of the highest transmission of $S_{21}$ = -10 dB at 2.9 GHz, we can infer a -5 dB (31\%) microwave-to-acoustic transduction efficiency for a single IDT.
The efficiency is mainly limited by the symmetric IDT design (maximum 50\% in one propagating direction) as well as by the mass loading and ohmic loss of aluminum. 
The fast variations of the $S_{21}$ spectrum is due to weak acoustic reflections between the IDT test pair.

The optical refractive index variations due to the acoustic wave (Fig.~\ref{Fig2}(c)) are calculated from a simulated strain profile and the photoelastic coefficients of LN \cite{Andrushchak2009JAP}. 
The refractive index variation in the $z$ direction ($\Delta n_z$) is slightly larger than that in other directions, and thus the optical TM mode is deflected more efficiency than the TE mode. 
We perform a three-dimensional finite-difference time-domain (FDTD) method simulation to investigate the spatial mode profile of the light that is deflected by the acoustic grating (Fig.~\ref{fig:FDTDsetup}). 
The FDTD simulation features actual device dimensions. 
Instead of a traveling acoustic wave, the acoustic grating in the FDTD simulation is represented by a steady refractive index profile. 
This approximation significantly reduces the computational expense without affecting the spatial mode profile of light. 
Absent the acoustic wave, the input light beam continues propagation, and no deflection is observed (Fig.~\ref{Fig2}(d)). 
With the acoustic wave, the input light beam is partially deflected (Fig.~\ref{Fig2}(e)). 
The deflected and the transmitted light beams are twice the Bragg angle apart in the far field. 
The width of the deflected beam is approximately given by
\begin{equation}
w_d \sim 2W\sin{\theta_B}+w_{in},
\label{Eq:wd}
\end{equation}
where $W$ is the width of the acoustic grating. 
For our device, with acoustic width of $W=100$ $\mu$m and $\sin{\theta_B}=0.3$, the width of the deflected beam is 60 $\mu$m larger than that of the input beam. 

\section{Experimental characterization of the LN AOFS}
\begin{figure*}
    \centering
    \includegraphics[center]{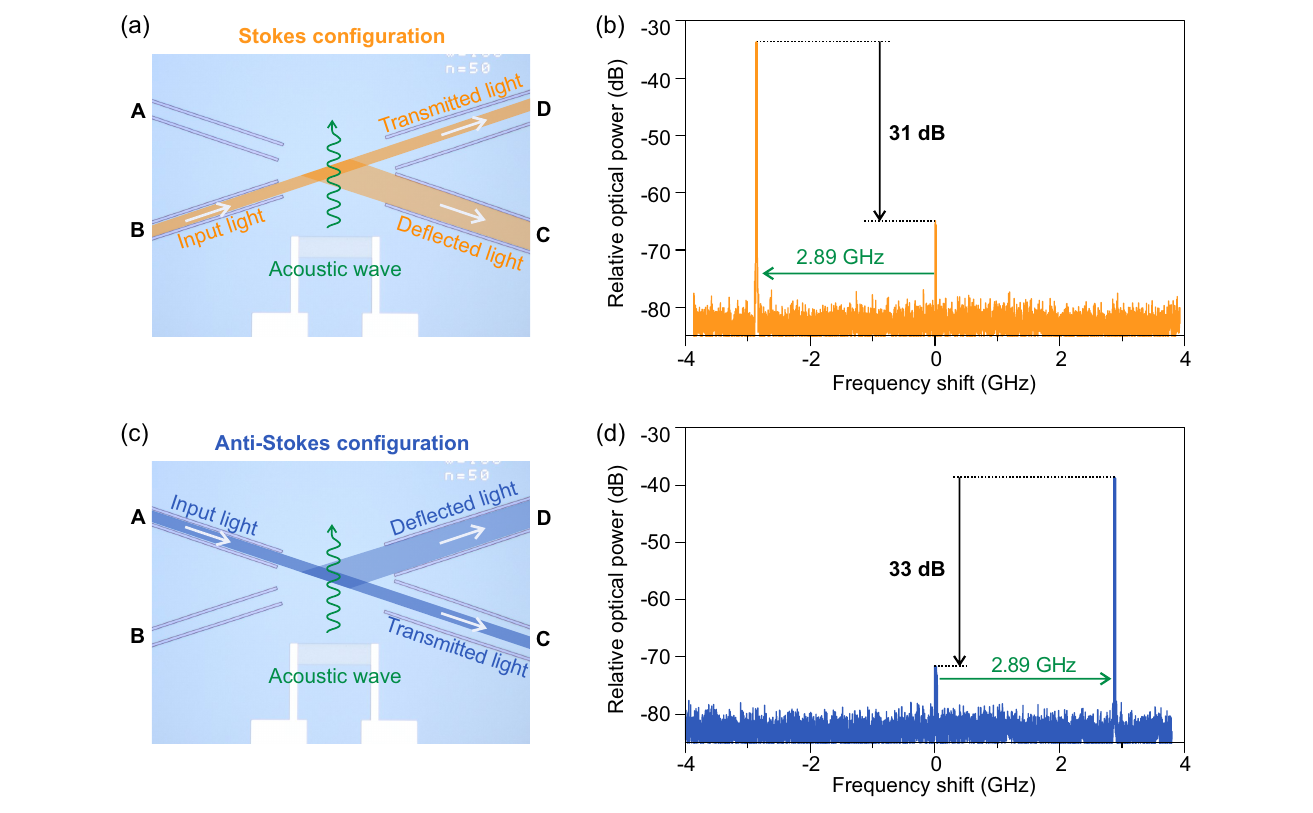}
    \caption{
    Measurements of the acousto-optic frequency shifting. 
    (a) Configuration for Stokes frequency shifting and (b) heterodyne measurement of frequency shifted light at Port C. 
    (c) Configuration for anti-Stokes frequency shifting and (d) heterodyne measurement of frequency shifted light at Port D.}
    \label{Fig3}
\end{figure*}

We experimentally characterize our LN AOFS in both the Stokes (Figs.~\ref{Fig3}(a) and \ref{Fig3}(b)) and anti-Stokes (Figs.~\ref{Fig3}(c) and \ref{Fig3}(d)) configurations.
We measure the optical spectra of the deflected output light by heterodyne detection using a second laser, which is red-detuned by a few gigahertz from the input laser. 
A schematic of the experimental setup is shown in Fig.\ref{fig:heterodyne}. 
The wavelength of the input light is 1597 nm, and the IDT is driven at 2.89 GHz with a microwave power of 15 dBm.
For the Stokes configuration, the deflected light is red-shifted by 2.89 GHz and exhibits a 31-dB carrier suppression (Fig.~\ref{Fig3}(b)). 
The carrier suppression of the deflected light is defined by the ratio of the optical powers at the shifted and carrier frequencies. 
For the anti-Stokes configuration, the deflected light is blue-shifted, and a 33-dB carrier suppression is observed (Fig.~\ref{Fig3}(d)). 
The presence of the carrier frequency light at the deflection output port is most likely caused by the optical scattering due to the on-chip structures, as it exists in absence of the acoustic wave. 
The on-to-off extinction ratio at the deflection port is also measured over to be 30 dB.
Any other undesired optical sideband is not observed above the noise floor in the heterodyne measurements, thus the sideband suppression rate exceeds 40 dB. 

We further characterize the efficiencies and bandwidths of our AOFS in the anti-Stokes configuration (Fig.~\ref{Fig4}). 
An insertion loss of -15 dB is determined by fiber-to-fiber optical transmission measurements through the transmitted light port. 
This loss is mainly due to the optical mode mismatch between the lensed fiber and waveguides and could be reduced by employing an on-chip coupler \cite{he2019ol}.
We define the on-chip frequency shifting efficiency as the ratio of optical powers between the deflected light and the transmitted light (when microwave input is off).
The measured efficiency linearly raises with microwave power and reaches 3.5\% at a 30-dBm microwave power applied to the IDT pads (Fig.~\ref{Fig4}(a)). 
Further increase of the microwave power can damage the IDT electrodes. 
We then measured the frequency-shifted optical power with varied microwave frequency (Fig.~\ref{Fig4}(b)).
We perform the homodyne detection of the deflected light (Fig.~\ref{fig:homodyne}). 
The 3-dB microwave bandwidth of our AOFS is 70 MHz for the Rayleigh mode (Mode I in Fig.~\ref{Fig4}(c)), which is determined by the IDT design. 
Although other acoustic modes have different frequencies, they feature a similar acoustic wavelength as defined by the IDT electrode pitch, and thus deflect the light at the same Bragg angle.
The optical bandwidth can be estimated as $\frac{\Lambda}{W}\lambda$ \cite{Saleh1991book}, which in our case gives a theoretically predicted bandwidth of 19 nm. The measured optical bandwidth of our AOFS is 14 nm, which is close to predicted value. The discrepancy could be attributed to the mode couplings in the wide waveguides.

\section{Acousto-optic frequency comb generation}

\begin{figure*}
    \centering
    \includegraphics[center]{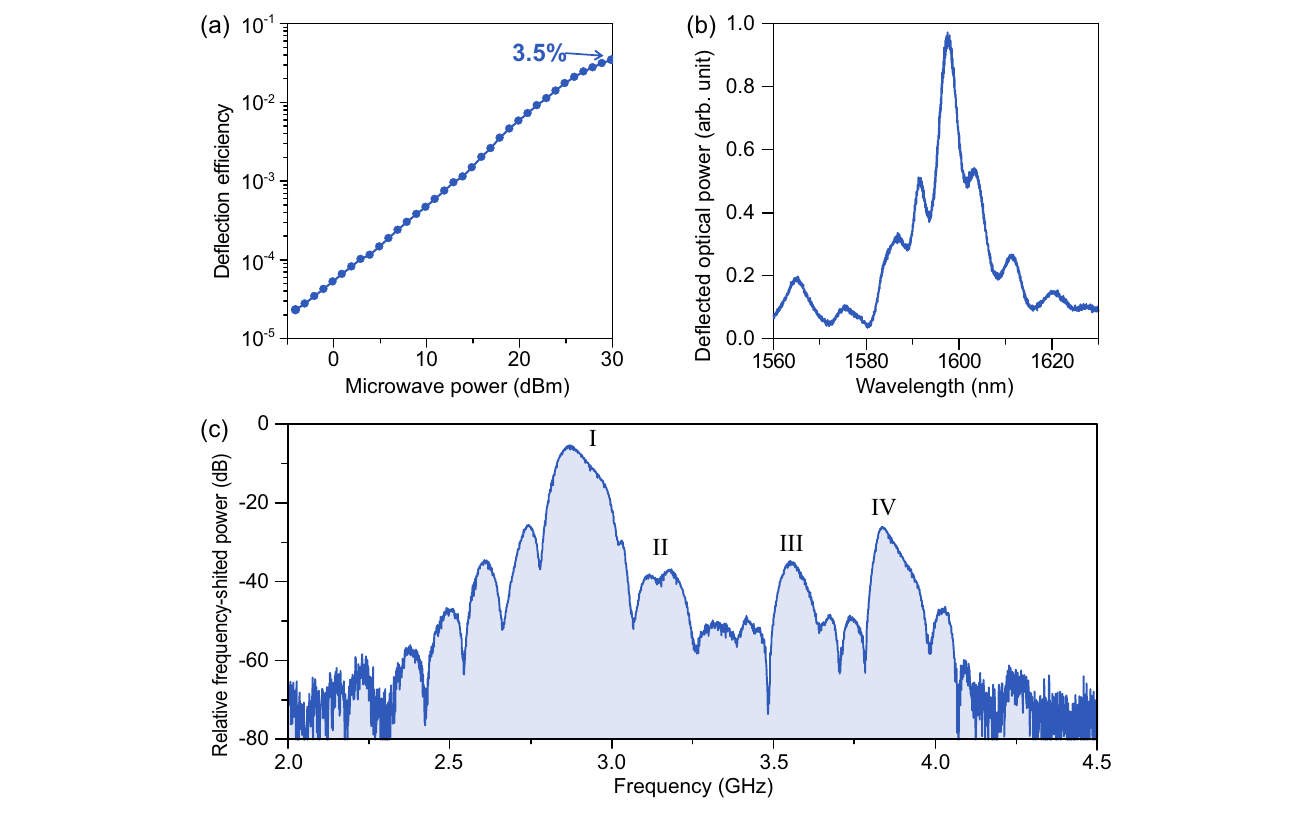}
    \caption{
    Characterization of the LN AOFS. 
    (a) Efficiency of the AOFS for varying microwave driving power. The acoustic frequency is 2.89 GHz.
    (b) Relative optical power of the frequency-shifted light for varying microwave frequency using the homodyne detection.
    For measurements in (a) and (b), the input optical wavelength is 1597 nm. 
    (c) Deflected optical power for different wavelengths of the input light. The acoustic frequency is 2.89 GHz. 
    Data in Figs.~\ref{Fig3} and \ref{Fig4} are measured from the same device. 
    }
    \label{Fig4}
\end{figure*}

\begin{figure*}[htb]
    \centering
    \includegraphics[center]{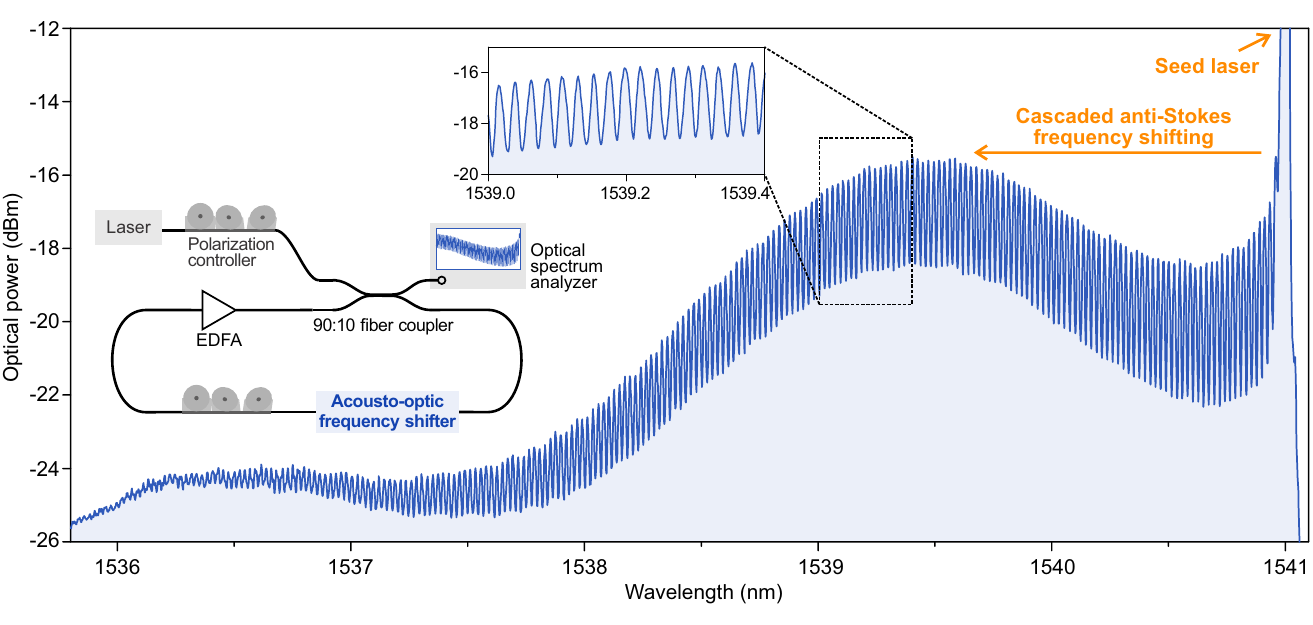}
    \caption{
    Optical frequency comb generation using an active frequency shifting loop.
    The wavelength of the seed laser is 1541 nm. 
    Inset: schematic of the setup for acousto-optic comb generation and magnification of the optical spectrum. 
    The optical spectrum analyzer features a resolution bandwidth of 0.02 nm (2.5 GHz at 1540 nm).
    }
    \label{Fig5}
\end{figure*}

An active acousto-optic frequency shifting loop, also known as frequency-shifted feedback laser \cite{Guillet2013PRA} can be used for optical frequency combs and pulse generation \cite{Duran2019PTL, Hale1990IEEEJQE, Kowalski1987apl}.
Moreover, this approach is of interest for various applications in optical real-time Fourier transformations \cite{Guillet2016Optica}, photonic microwave channeling \cite{Hao2018CLEO}, and optical frequency domain ranging \cite{Nakamura2000IEEE}.
For these applications, a broad frequency comb with gigahertz frequency spaced lines could improve bandwidth or ranging resolution.
Here, we generate an optical frequency comb with a 2.9 GHz line spacing using an active acousto-optic frequency shifting loop (Fig.~\ref{Fig5}).
The loop contains our AOFS and an erbium-doped fiber amplifier (EDFA) to compensate the optical loss (Fig.~\ref{Fig4} Inset). The loop is seeded with a laser at 1541 nm and couples light in and out by a 2$\times$2 90:10 fiber coupler.
The Bragg angle of the AOFS is chosen to be 17.45 degree to match the operating optical wavelength of our EDFA.
The generated frequency comb features over 200 comb lines spanning 0.6 THz (5 nm) optical bandwidth. 

The comb bandwidth is mainly limited by the gain saturation of the EDFA, in which the gain is reduced after the light has taken hundreds of round trips.
The demonstrated gigahertz acousto-optic frequency comb could find applications in dual-comb spectroscopy \cite{Duran2018OE}, wavelength-division-multiplexed communication, and on-chip short-pulse generation.

\section{Discussion and outlook}

We demonstrate an integrated gigahertz AOFS on a thin-film-LN-on-oxide substrate. 
The shift frequency could be tuned from hundreds of MHz to a few GHz by adjusting the Bragg angle.
Higher shift frequencies are possible for shorter optical wavelengths \cite{Desiatov2019Optica} or using materials with higher acoustic speed, such as aluminum nitride.
The microwave bandwidth could be extended by using multiple IDTs or chirped IDTs \cite{Fall2017JCSA}.
The optical wavelength of our AOFSs can be designed to any wavelength in the transparency window of LN, which is from visible to mid-infrared. 
For mid-infrared AOFSs, LN on sapphire substrates could be utilized to avoid optical absorption by the underlying oxide layer.
Although the efficiency of our AOFS is currently only 3.5\%, it could be further improved by employing unidirectional IDTs \cite{lehtonen2003IEEE} that could yield a 3 dB higher microwave-to-acoustic transduction efficiency than symmetric IDTs. 
Acoustic resonators \cite{shao2019PRApplied} could be employed to reduce the input microwave power for acousto-optic intensity modulators, in which the input light could be deflected by a resonant standing acoustic mode. 
A wider acoustic wave could also improve the deflection efficiency at the same total acoustic power \cite{li2019aplphoton}. 
However, as discussed in Eq.~\ref{Eq:wd}, widening the acoustic wave also broadening the deflected beam, which could be challenging for collecting the light back to a single mode waveguide, especially for large Bragg angles $\theta_B$ (high shift frequencies). 
On-chip lens \cite{Ren2014PTL} may be employed to better collect the broadened deflected light into a waveguide.
Moreover, our integrated acousto-optic platform, in which light is deflected by acoustic waves, would lead to the development of on-chip optical routers, isolators, and scanners \cite{Saleh1991book}, microwave spectral analyzers \cite{Turpin1981IEEE}, and carrier-envelope phase stabilizer \cite{Koke2010np}.

\section*{Appendix A: FDTD Simulation}
\begin{figure}[h]
\centering
\includegraphics[center]{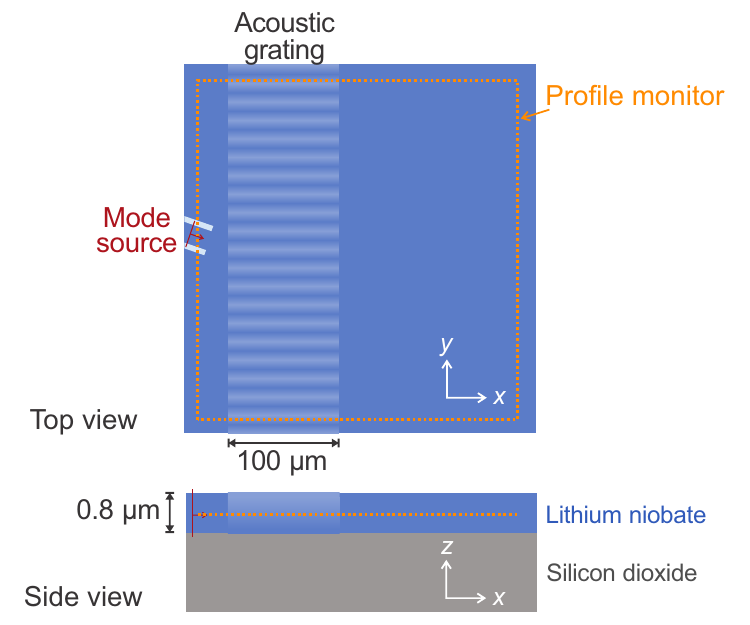}
\caption{Configuration of the FDTD simulation. 
The 0.8 $\mu$m lithium niobate layer is on silicon dioxide. 
The acoustic grating is simulated by a refractive index profile, which is derived from the strain profile of the acoustic wave.
The input waveguide is excited by a fundamental TM mode. 
}
\label{fig:FDTDsetup}
\end{figure}

To investigate the Bragg diffraction by the acoustic mode (grating), we perform a three-dimensional (3D) finite-difference time-domain (FDTD) method simulation using \textit{Lumerical FDTD Solutions}. 
We simulate the acousto-optic interaction of our device using its actual dimensions. 
The configuration of the simulation is illustrated in Fig.~\ref{fig:FDTDsetup}. 
The acoustic grating is simulated by a 3D refractive index profile. 
The steady refractive index profile used here significantly reduces the computational expense without affecting the spatial profiles.
The input waveguide is formed by two etched groves and is excited by the fundamental TM mode. 
The outer boundaries of the simulation region are set to the perfectly matched layers for the optical light.
The optical profile monitor at 0.4 $\mu$m from the top surface of the 0.8 $\mu$m lithium niobate layer is shown in Figs.~\ref{Fig2}(d) and \ref{Fig2}(e). 

\begin{figure}[h]
\centering
\includegraphics[center]{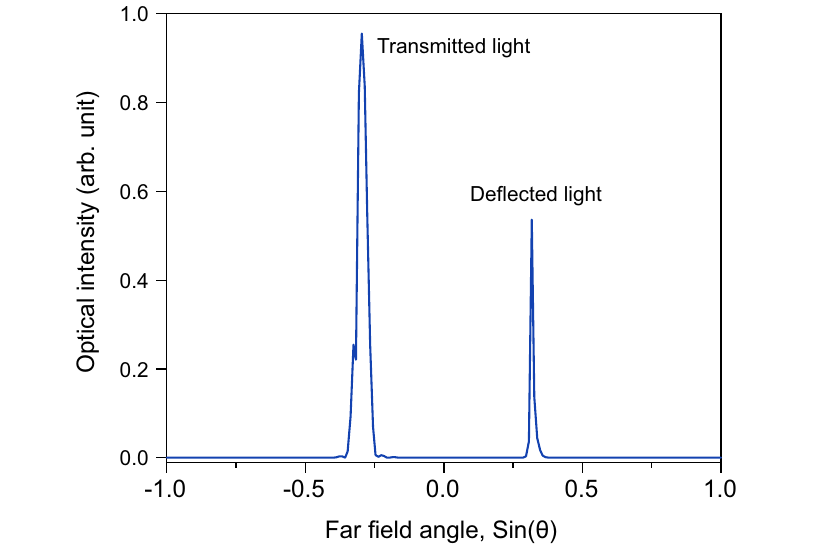}
\caption{Simulated far field pattern of the light after the acoustic grating. 
The deflected light propagates at the Bragg angle of $\sin{\theta_B} = 0.3$.}
\label{fig:farfield}
\end{figure}

The simulated far-field pattern after the acoustic grating (Fig.~\ref{fig:farfield}) shows the transmitted light at the input Bragg angle ($\sin{\theta} = -0.3$) of the acoustic grating and the deflected light at the opposite Bragg angle ($\sin{\theta} = 0.3$). 

\section*{Appendix B: Experimental setups}

\begin{figure*}[h!]
\centering
\includegraphics[center, width=\textwidth]{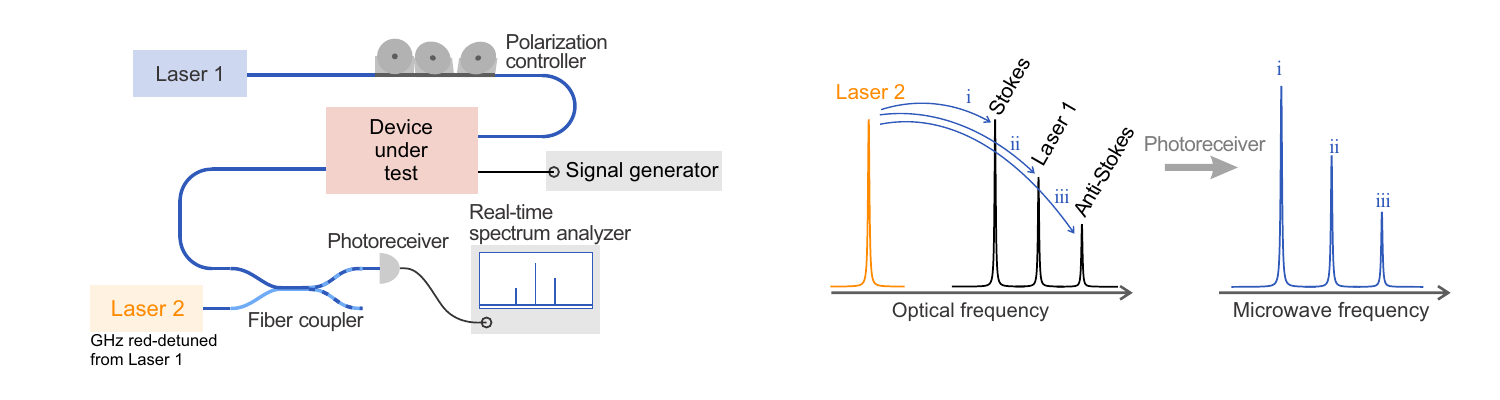}
\caption{The schematic diagrams of the heterodyne detection. 
The deflected output light from the device beats with a red-detuned Laser 2 on the photoreceiver and generates microwave signals corresponding to the carrier, the Stoke, and the anti-Stokes light.}
\label{fig:heterodyne}
\end{figure*}

\begin{figure*}[h!]
\centering
\includegraphics[center,width=\textwidth]{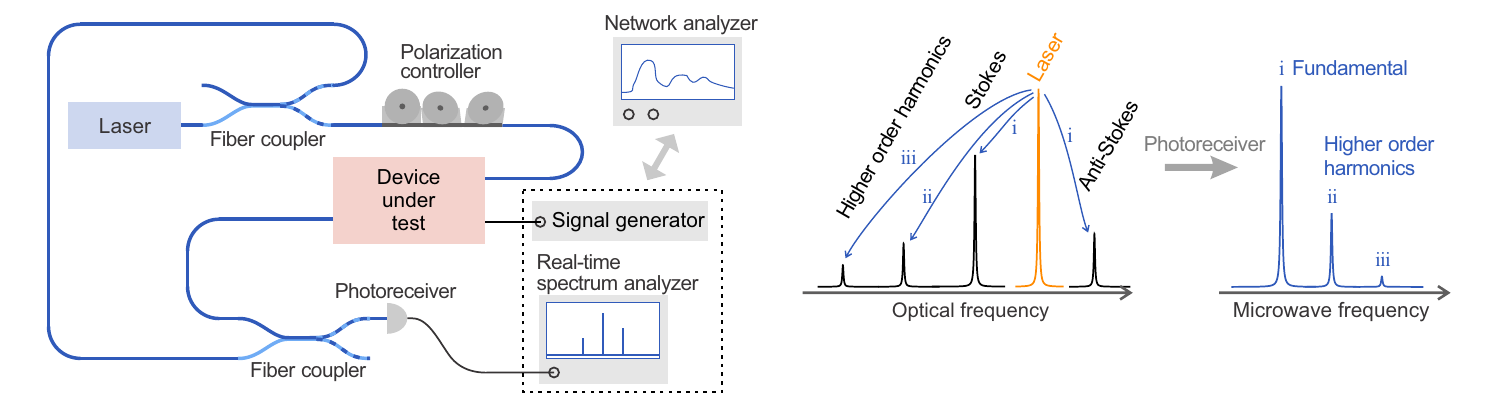}
\caption{The schematic diagrams of the homodyne detection. 
The deflected output light from the device beats with the input laser on the photoreceiver and generates microwave signals corresponding fundamental and higher order harmonic lights of the AOFS. 
The signal generator and the real-time spectrum analyzer are used for characterizing nonlinearity, and the network analyzer is used for characterizing the microwave bandwidth of the device. 
}
\label{fig:homodyne}
\end{figure*}

The schematic diagrams of the heterodyne and the homodyne detection setups are shown in Figs.~\ref{fig:heterodyne} and \ref{fig:homodyne}, respectively.

The heterodyne measurement uses two lasers and the Stokes and anti-Stokes frequency shifts are distinguished relative to the input carrier light. 
Here, the reference Laser 2 is red detuned from the input Laser 1.

The homodyne measurement uses the light from the same input laser in detection. This avoids frequency drifting between two lasers and reduces the noise floor on the spectrum analyzer by using narrow bandwidth, which increases the detection time.

\FloatBarrier

\section*{Funding}

National Science Foundation (NSF) (DMR-1231319, IIP-1827720);
Office of Naval Research (ONR) (N00014-15-1-2761);
Raytheon (A40210);
AQT Intelligent Quantum Networks and Technologies (INQNET);
Natural Sciences and Engineering Research Council of Canada;
U.S.~DOE (DE-SC0019219). 

\section*{Acknowledgments}
This work was performed in part at the Center for Nanoscale Systems (CNS), a member of the National Nanotechnology Coordinated Infrastructure Network (NNCI), which is supported by the National Science Foundation under NSF award no. 1541959. 
CNS is part of Harvard University.

\section*{Disclosures}
M.L.: HyperLight Corporation (I,C); Raytheon (F). \\
This document does not contain technology or Technical Data controlled under either the U.S.~International Traffic in Arms Regulations or the U.S.~Export Administration Regulations.

\bibliography{ref-LNAOFS}

\begin{thebibliography}{10}
\newcommand{\enquote}[1]{``#1''}

\bibitem{Eggleton2019NP}
B.~J. Eggleton, C.~G. Poulton, P.~T. Rakich, M.~J. Steel, and G.~Bahl,
  \enquote{Brillouin integrated photonics,} {\protect\JournalTitle{Nature
  Photonics}} \textbf{13}, 664--677 (2019).

\bibitem{cheng1992apl}
Z.~Y. Cheng and C.~S. Tsai, \enquote{Baseband integrated acousto‐optic
  frequency shifter,} {\protect\JournalTitle{Applied Physics Letters}}
  \textbf{60}, 12--14 (1992).

\bibitem{Dong2015adp}
C.~Dong, V.~Fiore, M.~C. Kuzyk, L.~Tian, and H.~Wang, \enquote{Optical
  wavelength conversion via optomechanical coupling in a silica resonator,}
  {\protect\JournalTitle{Annalen der Physik}} \textbf{527}, 100--106 (2015).

\bibitem{li2019aplphoton}
H.~Li, Q.~Liu, and M.~Li, \enquote{Electromechanical brillouin scattering in
  integrated planar photonics,} {\protect\JournalTitle{APL Photonics}}
  \textbf{4}, 080802 (2019).

\bibitem{Liu2019optica}
Q.~Liu, H.~Li, and M.~Li, \enquote{Electromechanical brillouin scattering in
  integrated optomechanical waveguides,} {\protect\JournalTitle{Optica}}
  \textbf{6}, 778--785 (2019).

\bibitem{salram2016np}
K.~C. Balram, M.~I. Davanco, J.~D. Song, and K.~Srinivasan, \enquote{Coherent
  coupling between radio frequency, optical, and acoustic waves in
  piezo-optomechanical circuits,} {\protect\JournalTitle{Nature Photonics}}
  \textbf{10}, 346--352 (2016).

\bibitem{Cai2019PR}
L.~Cai, A.~Mahmoud, M.~Khan, M.~Mahmoud, T.~Mukherjee, J.~Bain, and G.~Piazza,
  \enquote{Acousto-optical modulation of thin film lithium niobate waveguide
  devices,} {\protect\JournalTitle{Photonics Research}} \textbf{7}, 1003--1013
  (2019).

\bibitem{Jiang2019Optica}
W.~Jiang, R.~N. Patel, F.~M. Mayor, T.~P. McKenna, P.~Arrangoiz-Arriola, C.~J.
  Sarabalis, J.~D. Witmer, R.~Van~Laer, and A.~H. Safavi-Naeini,
  \enquote{Lithium niobate piezo-optomechanical crystals,}
  {\protect\JournalTitle{Optica}} \textbf{6}, 845--853 (2019).

\bibitem{Jiang2019arxiv}
W.~Jiang, C.~J. Sarabalis, Y.~D. Dahmani, R.~N. Patel, F.~M. Mayor, T.~P.
  McKenna, R.~Van~Laer, and A.~H. Safavi-Naeini, \enquote{Efficient
  bidirectional piezo-optomechanical transduction between microwave and optical
  frequency,} {\protect\JournalTitle{Nature Communications}} \textbf{11}, 1166
  (2020).

\bibitem{liang2017optica}
H.~Liang, R.~Luo, Y.~He, H.~Jiang, and Q.~Lin, \enquote{High-quality lithium
  niobate photonic crystal nanocavities,} {\protect\JournalTitle{Optica}}
  \textbf{4}, 1251 (2017).

\bibitem{shao2019optica}
L.~Shao, M.~Yu, S.~Maity, N.~Sinclair, L.~Zheng, C.~Chia, A.~Shams-Ansari,
  C.~Wang, M.~Zhang, K.~Lai, and M.~Lon{\v c}ar, \enquote{Microwave-to-optical
  conversion using lithium niobate thin-film acoustic resonators,}
  {\protect\JournalTitle{Optica}} \textbf{6}, 1498--1505 (2019).

\bibitem{marpaung2015optica}
D.~Marpaung, B.~Morrison, M.~Pagani, R.~Pant, D.-Y. Choi, B.~Luther-Davies,
  S.~J. Madden, and B.~J. Eggleton, \enquote{Low-power, chip-based stimulated
  brillouin scattering microwave photonic filter with ultrahigh selectivity,}
  {\protect\JournalTitle{Optica}} \textbf{2}, 76--83 (2015).

\bibitem{savchenkoc2011ol}
A.~A. Savchenkov, A.~B. Matsko, V.~S. Ilchenko, D.~Seidel, and L.~Maleki,
  \enquote{Surface acoustic wave opto-mechanical oscillator and frequency comb
  generator,} {\protect\JournalTitle{Optics Letters}} \textbf{36}, 3338--3340
  (2011).

\bibitem{Fan2019np}
L.~Fan, C.-L. Zou, N.~Zhu, and H.~X. Tang, \enquote{Spectrotemporal shaping of
  itinerant photons via distributed nanomechanics,}
  {\protect\JournalTitle{Nature Photonics}} \textbf{13}, 323--327 (2019).

\bibitem{Gundacarapu2019NP}
S.~Gundavarapu, G.~M. Brodnik, M.~Puckett, T.~Huffman, D.~Bose, R.~Behunin,
  J.~Wu, T.~Qiu, C.~Pinho, N.~Chauhan, J.~Nohava, P.~T. Rakich, K.~D. Nelson,
  M.~Salit, and D.~J. Blumenthal, \enquote{Sub-hertz fundamental linewidth
  photonic integrated brillouin laser,} {\protect\JournalTitle{Nature
  Photonics}} \textbf{13}, 60--67 (2019).

\bibitem{otterstrom2018science}
N.~T. Otterstrom, R.~O. Behunin, E.~A. Kittlaus, Z.~Wang, and P.~T. Rakich,
  \enquote{A silicon brillouin laser,} {\protect\JournalTitle{Science}}
  \textbf{360}, 1113--1116 (2018).

\bibitem{Fang2017NP}
K.~Fang, J.~Luo, A.~Metelmann, M.~H. Matheny, F.~Marquardt, A.~A. Clerk, and
  O.~Painter, \enquote{Generalized non-reciprocity in an optomechanical circuit
  via synthetic magnetism and reservoir engineering,}
  {\protect\JournalTitle{Nature Physics}} \textbf{13}, 465--471 (2017).

\bibitem{Kim2015NP}
J.~Kim, M.~C. Kuzyk, K.~Han, H.~Wang, and G.~Bahl, \enquote{Non-reciprocal
  brillouin scattering induced transparency,} {\protect\JournalTitle{Nature
  Physics}} \textbf{11}, 275--280 (2015).

\bibitem{Kittlaus2018NP}
E.~A. Kittlaus, N.~T. Otterstrom, P.~Kharel, S.~Gertler, and P.~T. Rakich,
  \enquote{Non-reciprocal interband brillouin modulation,}
  {\protect\JournalTitle{Nature Photonics}} \textbf{12}, 613--619 (2018).

\bibitem{Otterstrom2019Optica}
N.~T. Otterstrom, E.~A. Kittlaus, S.~Gertler, R.~O. Behunin, A.~L. Lentine, and
  P.~T. Rakich, \enquote{Resonantly enhanced nonreciprocal silicon brillouin
  amplifier,} {\protect\JournalTitle{Optica}} \textbf{6}, 1117--1123 (2019).

\bibitem{Kittlaus2020Arxiv}
E.~A. Kittlaus, W.~M. Jones, P.~T. Rakich, N.~T. Otterstrom, R.~E. Muller, and
  M.~Rais-Zadeh, \enquote{Electrically-driven acousto-optics and broadband
  non-reciprocity in silicon photonics,} {\protect\JournalTitle{arXiv
  e-prints}} p. 2004.01270 (2020).

\bibitem{Sohn2019arxiv}
D.~B. Sohn and G.~Bahl, \enquote{Direction reconfigurable nonreciprocal
  acousto-optic modulator on chip,} {\protect\JournalTitle{APL Photonics}}
  \textbf{4}, 126103 (2019).

\bibitem{sohn2018np}
D.~B. Sohn, S.~Kim, and G.~Bahl, \enquote{Time-reversal symmetry breaking with
  acoustic pumping of nanophotonic circuits,} {\protect\JournalTitle{Nature
  Photonics}} \textbf{12}, 91--97 (2018).

\bibitem{savage2010np}
N.~Savage, \enquote{Acousto-optic devices,} {\protect\JournalTitle{Nature
  Photonics}} \textbf{4}, 728--729 (2010).

\bibitem{Higuma2001EL}
K.~Higuma, S.~Oikawa, Y.~Hashimoto, H.~Nagata, and M.~Izutsu, \enquote{X-cut
  lithium niobate optical single-sideband modulator,}
  {\protect\JournalTitle{Electronics Letters}} \textbf{37}, 515--516 (2001).

\bibitem{ogiso2010IEEEptl}
Y.~Ogiso, Y.~Tsuchiya, S.~Shinada, S.~Nakajima, T.~Kawanishi, and H.~Nakajima,
  \enquote{High extinction-ratio integrated mach–zehnder modulator with
  active y-branch for optical ssb signal generation,}
  {\protect\JournalTitle{IEEE Photonics Technology Letters}} \textbf{22},
  941--943 (2010).

\bibitem{Houtz2009OE}
R.~Houtz, C.~Chan, and H.~M{\" u}ller, \enquote{Wideband, efficient optical
  serrodyne frequency shifting with a phase modulator and a nonlinear
  transmission line,} {\protect\JournalTitle{Optics Express}} \textbf{17},
  19235--19240 (2009).

\bibitem{Johnson2010OL}
D.~M.~S. Johnson, J.~M. Hogan, S.~w. Chiow, and M.~A. Kasevich,
  \enquote{Broadband optical serrodyne frequency shifting,}
  {\protect\JournalTitle{Optics Letters}} \textbf{35}, 745--747 (2010).

\bibitem{Spuesens2016CLEO}
T.~Spuesens, Y.~Li, P.~Verheyen, G.~Lepage, S.~Balakrishnan, P.~Absil, and
  R.~Baets, \enquote{Integrated optical frequency shifter on a silicon
  platform,} in \emph{Conference on Lasers and Electro-Optics,}  (Optical
  Society of America, 2016), OSA Technical Digest (2016), p. SF2G.1.

\bibitem{Preble2007NPhoton}
S.~F. Preble, Q.~Xu, and M.~Lipson, \enquote{Changing the colour of light in a
  silicon resonator,} {\protect\JournalTitle{Nature Photonics}} \textbf{1},
  293--296 (2007).

\bibitem{savchenkov2009ol}
A.~A. Savchenkov, W.~Liang, A.~B. Matsko, V.~S. Ilchenko, D.~Seidel, and
  L.~Maleki, \enquote{Tunable optical single-sideband modulator with complete
  sideband suppression,} {\protect\JournalTitle{Optics Letters}} \textbf{34},
  1300--1302 (2009).

\bibitem{Yu2018PR}
B.-M. Yu, J.-M. Lee, C.~Mai, S.~Lischke, L.~Zimmermann, and W.-Y. Choi,
  \enquote{Single-chip si optical single-sideband modulator,}
  {\protect\JournalTitle{Photonics Research}} \textbf{6}, 6--11 (2018).

\bibitem{Andrushchak2009JAP}
A.~S. Andrushchak, B.~G. Mytsyk, H.~P. Laba, O.~V. Yurkevych, I.~M. Solskii,
  A.~V. Kityk, and B.~Sahraoui, \enquote{Complete sets of elastic constants and
  photoelastic coefficients of pure and mgo-doped lithium niobate crystals at
  room temperature,} {\protect\JournalTitle{Journal of Applied Physics}}
  \textbf{106}, 073510 (2009).

\bibitem{he2019ol}
L.~He, M.~Zhang, A.~Shams-Ansari, R.~Zhu, C.~Wang, and M.~Lon{\v c}ar,
  \enquote{Low-loss fiber-to-chip interface for lithium niobate photonic
  integrated circuits,} {\protect\JournalTitle{Optics Letters}} \textbf{44},
  2314--2317 (2019).

\bibitem{Saleh1991book}
B.~E. Saleh and M.~C. Teich, \emph{Ch. 20 Acousto-optics} (John Wiley \& Sons,
  New York, 1991).

\bibitem{Guillet2013PRA}
H.~Guillet~de Chatellus, E.~Lacot, W.~Glastre, O.~Jacquin, and O.~Hugon,
  \enquote{Theory of talbot lasers,} {\protect\JournalTitle{Physical Review A}}
  \textbf{88}, 033828 (2013).

\bibitem{Duran2019PTL}
V.~Dur{\'a}n, H.~G.~d. Chatellus, C.~Schn{\'e}belin, K.~Nithyanandan,
  L.~Djevarhidjian, J.~Clement, and C.~R. Fern{\'a}ndez-Pousa, \enquote{Optical
  frequency combs generated by acousto-optic frequency-shifting loops,}
  {\protect\JournalTitle{IEEE Photonics Technology Letters}} \textbf{31},
  1878--1881 (2019).

\bibitem{Hale1990IEEEJQE}
P.~D. Hale and F.~V. Kowalski, \enquote{Output characterization of a frequency
  shifted feedback laser: theory and experiment,} {\protect\JournalTitle{IEEE
  Journal of Quantum Electronics}} \textbf{26}, 1845--1851 (1990).

\bibitem{Kowalski1987apl}
F.~V. Kowalski, J.~A. Squier, and J.~T. Pinckney, \enquote{Pulse generation
  with an acousto‐optic frequency shifter in a passive cavity,}
  {\protect\JournalTitle{Applied Physics Letters}} \textbf{50}, 711--713
  (1987).

\bibitem{Guillet2016Optica}
H.~Guillet~de Chatellus, L.~R. Cort{\'e}s, and J.~Aza{\~n}a, \enquote{Optical
  real-time fourier transformation with kilohertz resolutions,}
  {\protect\JournalTitle{Optica}} \textbf{3}, 1--8 (2016).

\bibitem{Hao2018CLEO}
W.~Hao, Y.~Dai, F.~Yin, Y.~Zhou, J.~Li, and K.~Xu, \enquote{Photonic microwave
  channelization based on frequency shifted feedback laser and delayed coherent
  detection,} in \emph{Conference on Lasers and Electro-Optics,}  (Optical
  Society of America, 2018), OSA Technical Digest (online), p. JW2A.73.

\bibitem{Nakamura2000IEEE}
K.~Nakamura, T.~Hara, M.~Yoshida, T.~Miyahara, and H.~Ito, \enquote{Optical
  frequency domain ranging by a frequency-shifted feedback laser,}
  {\protect\JournalTitle{IEEE Journal of Quantum Electronics}} \textbf{36},
  305--316 (2000).

\bibitem{Duran2018OE}
V.~Dur{\'a}n, C.~Schn{\'e}belin, and H.~Guillet~de Chatellus, \enquote{Coherent
  multi-heterodyne spectroscopy using acousto-optic frequency combs,}
  {\protect\JournalTitle{Optics Express}} \textbf{26}, 13800--13809 (2018).

\bibitem{Desiatov2019Optica}
B.~Desiatov, A.~Shams-Ansari, M.~Zhang, C.~Wang, and M.~Lon{\v c}ar,
  \enquote{Ultra-low-loss integrated visible photonics using thin-film lithium
  niobate,} {\protect\JournalTitle{Optica}} \textbf{6}, 380--384 (2019).

\bibitem{Fall2017JCSA}
D.~Fall, M.~Duquennoy, M.~Ouaftouh, N.~Smagin, B.~Piwakowski, and F.~Jenot,
  \enquote{Generation of broadband surface acoustic waves using a dual
  temporal-spatial chirp method,} {\protect\JournalTitle{The Journal of the
  Acoustical Society of America}} \textbf{142}, EL108--EL112 (2017).

\bibitem{lehtonen2003IEEE}
S.~Lehtonen, V.~P. Plessky, C.~S. Hartmann, and M.~M. Salomaa,
  \enquote{Unidirectional saw transducer for gigahertz frequencies,}
  {\protect\JournalTitle{IEEE Transactions on Ultrasonics, Ferroelectrics, and
  Frequency Control}} \textbf{50}, 1404--1406 (2003).

\bibitem{shao2019PRApplied}
L.~Shao, S.~Maity, L.~Zheng, L.~Wu, A.~Shams-Ansari, Y.-I. Sohn, E.~Puma, M.~N.
  Gadalla, M.~Zhang, C.~Wang, E.~Hu, K.~Lai, and M.~Lon{\v c}ar,
  \enquote{Phononic band structure engineering for high-q gigahertz surface
  acoustic wave resonators on lithium niobate,} {\protect\JournalTitle{Physical
  Review Applied}} \textbf{12}, 014022 (2019).

\bibitem{Ren2014PTL}
G.~Ren, T.~G. Nguyen, and A.~Mitchell, \enquote{Gaussian beams on a
  silicon-on-insulator chip using integrated optical lenses,}
  {\protect\JournalTitle{IEEE Photonics Technology Letters}} \textbf{26},
  1438--1441 (2014).

\bibitem{Turpin1981IEEE}
T.~M. Turpin, \enquote{Spectrum analysis using optical processing,}
  {\protect\JournalTitle{Proceedings of the IEEE}} \textbf{69}, 79--92 (1981).

\bibitem{Koke2010np}
S.~Koke, C.~Grebing, H.~Frei, A.~Anderson, A.~Assion, and G.~Steinmeyer,
  \enquote{Direct frequency comb synthesis with arbitrary offset and
  shot-noise-limited phase noise,} {\protect\JournalTitle{Nature Photonics}}
  \textbf{4}, 462--465 (2010).

\end{thebibliography}

\end{document}